\documentclass[twocolumn,aps,prd]{revtex4-1}

\pdfoutput=1
\usepackage[pdftex]{graphicx}
\usepackage[unicode=true, bookmarks=false,bookmarksnumbered=true,bookmarksopen=true, breaklinks=false,backref=false,pagebackref=false, colorlinks=true]{hyperref}
\hypersetup{pdftitle={ },pdfauthor={ },linkcolor=black,citecolor=black, urlcolor=black, filecolor=black,pdfpagelayout=OneColumn,pdfnewwindow=true,pdfstartview=XYZ, plainpages=false}
\usepackage{color}

\usepackage{braket}
\usepackage[T1]{fontenc}
\usepackage{amsmath}
\usepackage{amsthm}
\usepackage{amssymb}
\usepackage{url}
\usepackage[english]{babel}
\usepackage{dcolumn}
\usepackage{bm}
\usepackage{subfig}
\usepackage{float}
\usepackage{url} 

\newcommand{\Tr}{\mbox{\rm Tr\,}}
\newcommand{\ReC}{\mbox{\rm Re\,}}

\newcommand{\be}{\begin{equation}}
\newcommand{\ee}{\end{equation}}
\newcommand{\bea}{\begin{eqnarray}}
\newcommand{\eea}{\end{eqnarray}}

\newcommand{\bit}{\begin{itemize}}
\newcommand{\eit}{\end{itemize}}

\graphicspath{{figures/}}
\usepackage{verbatim}

\begin{document}

\title{Lattice QCD computation of the SU(3) String Tension critical curve}
\author{N. Cardoso}
\email{nunocardoso@cftp.ist.utl.pt}
\author{P. Bicudo}
\email{bicudo@ist.utl.pt}
\affiliation{CFTP, Departamento de F\'{\i}sica, Instituto Superior T\'{e}cnico
(Universidade T\'{e}cnica de Lisboa),
Av. Rovisco Pais, 1049-001 Lisboa, Portugal}

\begin{abstract}

We investigate the critical curve of the string tension $\sigma(T)$ as a function of temperature in quenched gauge invariant SU(3) lattice gauge theory.
We extract $\sigma(T)$ from the colour averaged free energy of a static quark-antiquark pair.
To compute the free energy, we utilize a pair of gauge invariant Polyakov loop and antiloop correlations, and apply the multi-hit procedure to enhance the signal to noise ratio.
We find that the string tension departs from the  zero temperature $\sigma_0$ at $T \simeq 0.5 T_c$.
We cover the relevant temperature range from $0.5T_c$ up to the confinement temperature $T_c$ using 54 different sets of pure gauge lattice configurations with four temporal extensions (4,6,8,12), different $\beta$ and a spatial volume of $48^3$ in lattice units.

\end{abstract}
\maketitle

\section{Introduction}

The string tension $\sigma(T)$ is a relevant order parameter to study confinement. 
While above the deconfinement temperature $T_c$ the simplest order parameter is the Polyakov loop, below $T_c$ the expectation value of a single Polyakov loop vanishes.  Below $T_c$, to study the decrease of confinement with $T$ a new order parameter must be used, and we utilize here the string tension. 
The string tension $\sigma(t)$ parameterizes the confining sector of the quark-antiquark potential, which increases linearly with distance. 
At finite temperature the quark-antiquark potential $V(r)$ is extended to the quark-antiquark free energy $F(r,T)$.

Moreover we are interested in the string tension and in the confining quark-antiquark free energy, 
since it leads to chiral symmetry breaking
\cite{Bicudo:2010hg}. 
It also dominates the hadron spectrum. 
To study chiral symmetry and the hadron spectrum at finite $T$, 
It is important to know the string tension $\sigma(T)$ behaviour at all temperatures. 

The existing lattice QCD results for the string tension critical curve have been computed by the Bielefeld group
\cite{Kaczmarek:1999mm} ,
who have studied in detail the heavy quark potentials in the confined and deconfined phases at finite temperature.  
In the confined phase, the Bielefeld Group presented results for the string tension in the region $[0.8 T_c,T_c[$ with lattice size of $32^3\times 4$ generated with a tree level Symanzik-improved gauge action.  Their results confirmed a first order deconfinement phase transition, as expected for SU(3).
Bialas et al., 
\cite{Bialas:2009pt} 
have studied the string tension behaviour at finite temperature in three dimensional SU(3) gauge theory . 
Bicudo 
\cite{Bicudo:2010hg}
compared the string tension points obtained by Bielefeld group 
\cite{Kaczmarek:1999mm}
with different order parameter curves 
and found empirically that the ferromagnetic critical curve is the one closer to the Bielefeld data.

This works continues the study of the SU(3) string tension, first computed by the Bielefeld Group 
\cite{Kaczmarek:1999mm}, and we utilize a similar technique of colour averaged free energy. 
We study a wider range of temperatures in order to map the critical curve of $\sigma(T)$.
In section II, we present in detail our method to extract the string tension at finite temperature.
in section III we present and discuss our results and in section IV we conclude.

\section{Our SU(3) lattice QCD framework}

We compute $\sigma(T)$ with in quenched SU(3) lattice QCD,  fitting the linear confinement from the colour averaged free energy of a quark-antiquark pair.
The colour averaged free energy of a static quark-antiquark pair is computed with the correlation of a pair of Polyakov loop and antiloop correlations,
\begin{equation}
   \exp\left( - \frac{F_{q\overline{q}} (R,T)}{T}  \right) = \Braket{ L^\dagger(\vec{x}) L(\vec{y})  } \ ,
   \label{favg}
\end{equation}
where $R=|\vec{x}-\vec{y}|$, $T=1/(a N_t)$ and the temperature $T$, in units of the Boltzmann  constant $K=1$, is the inverse of the temporal extent of the lattice,
\begin{equation}
  T={1 \over a N_t} \ , 
 \label{temp}
\end{equation}
$a$ is the lattice spacing and $N_t$ is the number of links in the temporal direction.

A Polyakov loop is a Wilson line of lattice link in the temporal direction and closed with the periodic boundary condition of the lattice,
\begin{equation}
   L(\vec{x}) = \frac{1}{3} \ReC \Tr \prod_{\tau=0}^{N_t-1} U_3(\vec{x},\tau)        
   \label{polyakov}
\end{equation}
where $U_\mu(\vec{x},\tau) $ with $\mu=3$ is the time direction link. 
The Polyakov loop is an order parameter for the deconfinement transition, i.e.  
$\Braket{L}=0$ in the confined phase for $T<T_c$ and $\Braket{L}>0$ in the deconfined phase.
Since a single Polyakov loop vanishes in the confined phase, 
to measure confinement we need a higher number of Polyakov loops. Here utilize a pair of Polyakov loop-antiloop.

\begin{figure}[t!]
\begin{center}
    \includegraphics[width=6.5cm]{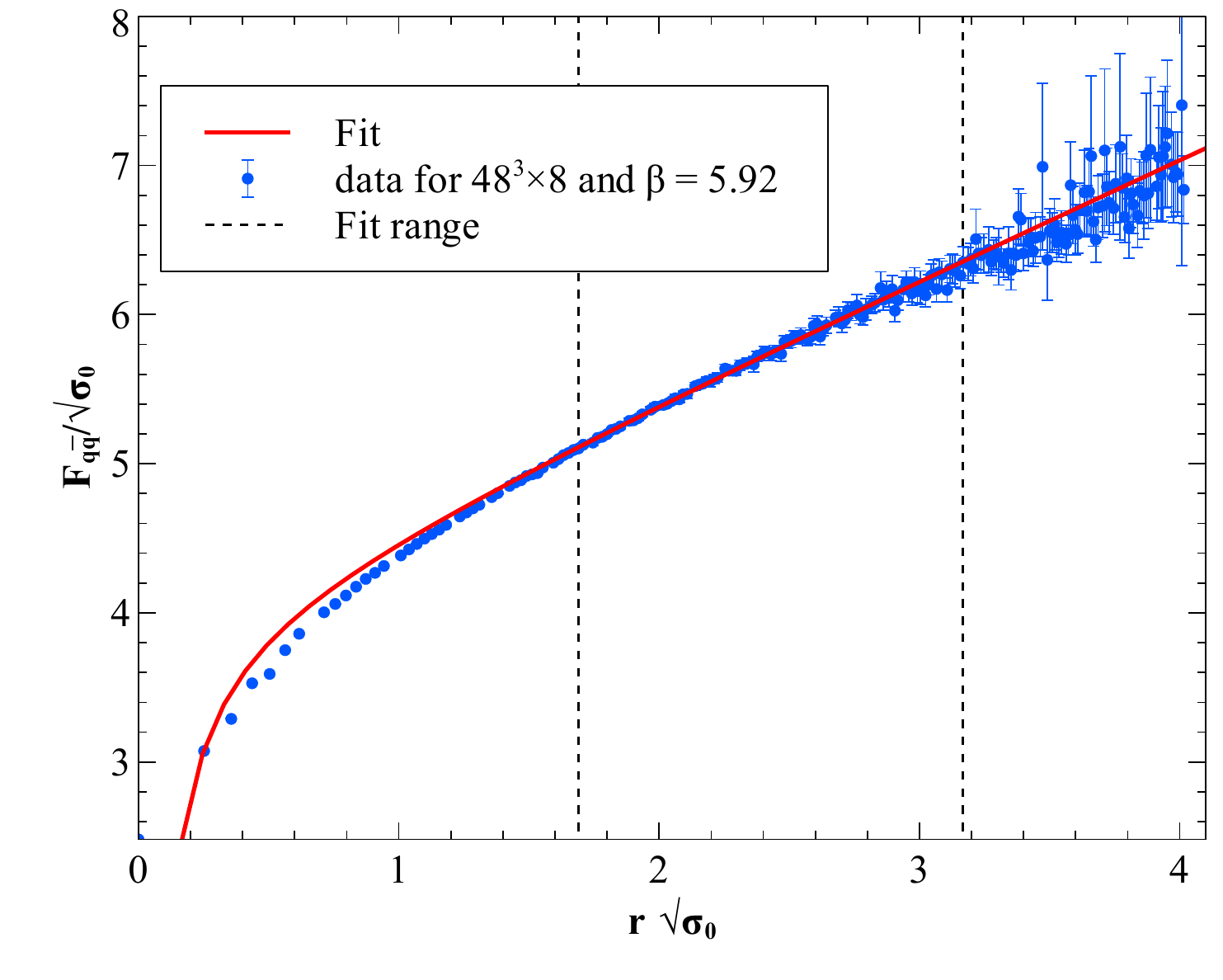}
    \caption{Colour averaged free energy results for $N_t=8$, $L/a=48$ and $\beta=5.92$, at $T=0.788T_c$. The solid line corresponds to the fit,  $a_0+a_1/r+a_2 r$ with $a_1=-\frac{\pi}{12}$, in the range defined by the dashed lines.}
    \label{fig:Fit_48_3_8_b5.92}
\end{center}
\end{figure}

The temperature and distance in Eq.  (\ref{favg}) can be made dimensionless using the $\sigma_0 = \sigma(0)$
string tension at zero temperature , 
\begin{equation}
T^*=\frac{1}{N_\tau\, a\sqrt\sigma_0} = \frac{T}{\sqrt\sigma_0} \ .
\end{equation}
and 
\begin{equation}
 r = R\, a\sqrt\sigma_0 \ .
\end{equation}
We fit the free energy $F_{q\overline{q}}(r,T^*)$ with
\begin{equation}
 a_0+ \frac{a_1}{r}+a_2 r,
   \label{ffit}
\end{equation}
where $a_1$ is fixed to the L\"{u}scher Coulomb $ \pi / 12$ term and $a_2$ provides $\sigma(T)/\sigma_0$.
We also tried a fit with a logarithmic term, however the fits were not stable, and thus we abandoned any logarithmic term.

To improve the signal in the Polyakov loop correlation functions and reduce the error, we employ the multi-hit procedure for the time links, \cite{Brower:1981vt,Parisi:1983hm}. The SU(3) temporal links can be integrated out analytically and substituted by their average value,
\begin{equation}
 U_3(\mathbf{x})\rightarrow \overline{U}_3(\mathbf{x})\equiv \frac{\int dU U_3(\mathbf{x})\,e^{\frac{1}{3}\beta \Tr\left(U_3(\mathbf{x})F^\dagger(\mathbf{x})\right)}}{\int dU\,e^{\frac{1}{3}\beta \Tr\left(U_3(\mathbf{x})F^\dagger(\mathbf{x})\right)}}\ ,
\end{equation}
where $F(\mathbf{x})$ is the staple attached to a specific time link. We applied the multi-hit numerically to the time links using
\begin{equation}
 U_3(\mathbf{x})\rightarrow \overline{U}_3(\mathbf{x}) = \frac{1}{n}\sum_{i=1}^n U_3^{(i)}\ ,
\end{equation}
with $U_3^{(i)}$ generated with the pseudo-heatbath algorithm and maintaining all the neighbouring links fixed.

\begin{figure}[t!]
\begin{center}
    \includegraphics[width=6.5cm]{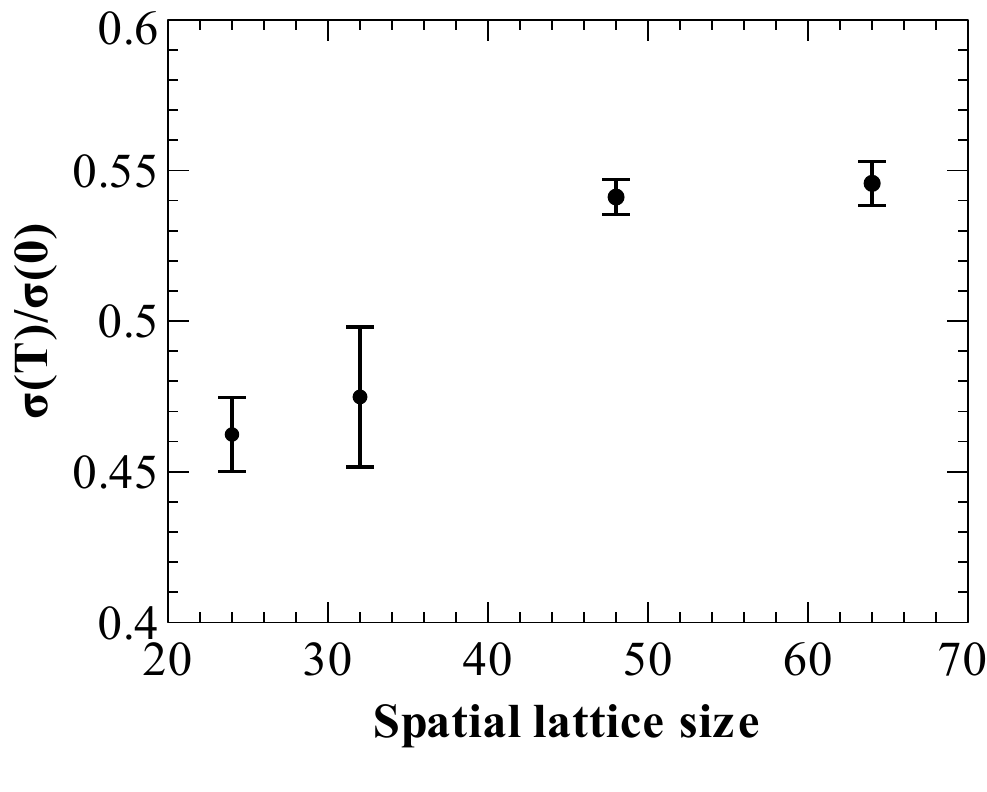}
    \caption{Large volume study of the string tension. We study $\sigma$ for spatial lattice volumes ($24^3,32^3,48^3$ and $64^3$) for $N_t=4$ and $T=0.95T_c$.}
    \label{stft_nt4_b5}
\end{center}
\end{figure}

To scan the temperatures, determined by Eq. (\ref{temp})
we utilize different $\beta$ and $N_t$, because the lattice spacing $a$ is a function of $\beta$.
To set the scale of the lattice spacing $a$ in physical units we use the equations fitted in SU(3) lattice QCD
by Edwards et al.  
\cite{Edwards:1997xf},
\begin{eqnarray}
\left(a\sqrt\sigma_0\right)(g) & = & f(g^2) \left[ 1 + b_1 \,\hat{a}(g)^2 + b_2 \,\hat{a}(g)^4 +\right.\nonumber \\
 &  & \left. + b_3 \,\hat{a}(g)^6 \right]/b_0 
\end{eqnarray}
where $g$ is the coupling constant of the Wilson action,
and  where Edwards et al.  
\cite{Edwards:1997xf}
obtained $b_0 = 0.01364$, $b_1 = 0.2731$, $b_2 = -0.01545$ and $b_3 = 0.01975$, valid in the region $5.6 \leq \beta \leq 6.5$,
\begin{equation}
\hat{a}(g) = \frac{f(g^2)}{f(g^2(\beta=6.0))},
\end{equation}
\begin{equation}
f(g^2) = \left(b_0g^2\right)^{-\frac{b_1}{2b_0^2}} \exp\left( -\frac{1}{2b_0g^2}  \right),
\end{equation}
and
\begin{equation}
b_0 = \frac{11}{(4\pi)^2},\quad\quad b_1 = \frac{102}{(4\pi)^4} \ .
\end{equation}

Finally, the normalized temperature, $T/T_c$, is given by
\begin{equation}
\frac{T}{T_c}=\frac{\left(a\sqrt\sigma_0\right)(g(\beta_c))}{\left(a\sqrt\sigma_0\right)(g(\beta))}
\end{equation}
where $\beta_c$ was obtained by Lucini et al. \cite{Lucini:2003zr}.
Thus we determine the temperature $T/T_c$ from $\beta$ and $N_t$.

\begin{table}
\caption{Summary of parameters and results for the string tension at finite $T$  
with $48^3 \times Nt$ lattices. For the $\beta$ marked with an asterisk *, corresponding to $ 0.95 T_c \leq T < T_c$, at finite wolume, we find a contamination with some deconfined configurations. We removed the wrong configurations from our average. 
}
\label{tab:stringtension}
\begin{ruledtabular}
\begin{tabular}{cc|cccc}
 $\ \ \ \beta \ \ \ $ & $ \ N_t \ $ &  $ \ T/T_c \ $ & $a\sqrt{\sigma_0}$  & $ \ \ \ \sigma(T)/\sigma_0 \ \ \ $ & $\#$ config.\\
\hline
 5.69* 	&	 4 	&	 0.994 	&	 0.4011 	&	 0.3468(15)		&	 223\\
 5.68 	&	 4 	&	 0.971 	&	 0.4108 	&	 0.4728(77)		&	 225\\
 5.67 	&	 4 	&	 0.948 	&	 0.4208 	&	 0.5412(57)		&	 450\\
 5.66 	&	 4 	&	 0.925 	&	 0.4312 	&	 0.5879(42)		&	 225\\
 5.65 	&	 4 	&	 0.902 	&	 0.4421 	&	 0.6148(62)		&	 225\\
 5.64 	&	 4 	&	 0.879 	&	 0.4535 	&	 0.6676(179)		&	 225\\
 5.63 	&	 4 	&	 0.857 	&	 0.4654 	&	 0.7295(128)		&	 225\\
 5.62 	&	 4 	&	 0.835 	&	 0.4778 	&	 0.7439(78)		&	 300\\
 5.61 	&	 4 	&	 0.813 	&	 0.4908 	&	 0.7577(49)		&	 325\\
\hline											
 5.89$^*$ 	&	 6 	&	 0.993 	&	 0.2659 	&	 0.3418(28)		&	 206\\
 5.87 	&	 6 	&	 0.957 	&	 0.2759 	&	 0.4408(60)		&	 224\\
 5.80 	&	 6 	&	 0.836 	&	 0.3159 	&	 0.7754(39)		&	 225\\
 5.79 	&	 6 	&	 0.819 	&	 0.3223 	&	 0.7740(64)		&	 225\\
 5.78 	&	 6 	&	 0.802 	&	 0.3290 	&	 0.8071(149)		&	 225\\
\hline
 6.062$^*$ 	&	 8 	&	 0.999 	&	 0.1987 	&	 0.4276(63)		&	 98\\
 6.061$^*$ 	&	 8 	&	 0.998 	&	 0.1990 	&	 0.3865(159)		&	 93\\
 6.06$^*$ 	&	 8 	&	 0.996 	&	 0.1993 	&	 0.3763(134)		&	 127\\
 6.055$^*$ 	&	 8 	&	 0.988 	&	 0.2009 	&	 00.4041(87)		&	 175\\
 6.05$^*$ 	&	 8 	&	 0.980 	&	 0.2025 	&	 0.4354(68)		&	 199\\
 6.04$^*$ 	&	 8 	&	 0.965 	&	 0.2058 	&	 0.5003(67)		&	 238\\
 6.03 	&	 8 	&	 0.949 	&	 0.2092 	&	 0.5336(92)		&	 225\\
 6.02 	&	 8 	&	 0.934 	&	 0.2126 	&	 0.5783(84)		&	 225\\
 6.01 	&	 8 	&	 0.919 	&	 0.2161 	&	 0.5950(47)		&	 225\\
 6.00 	&	 8 	&	 0.904 	&	 0.2197 	&	 0.6526(108)		&	 228\\
 5.99 	&	 8 	&	 0.889 	&	 0.2234 	&	 0.6974(64)		&	 225\\
 5.98 	&	 8 	&	 0.874 	&	 0.2272 	&	 0.7075(33)		&	 227\\
 5.97 	&	 8 	&	 0.859 	&	 0.2311 	&	 0.7281(205)		&	 269\\
 5.96 	&	 8 	&	 0.845 	&	 0.2350 	&	 0.7395(52)		&	 225\\
 5.94 	&	 8 	&	 0.816 	&	 0.2433 	&	 0.7683(61)		&	 225\\
 5.92 	&	 8 	&	 0.788 	&	 0.2520 	&	 0.7943(57)		&	 275\\
 5.90 	&	 8 	&	 0.760 	&	 0.2611 	&	 0.8400(130)		&	 225\\
 5.89 	&	 8 	&	 0.747 	&	 0.2659 	&	 0.8674(81)		&	 227\\
 5.87 	&	 8 	&	 0.720 	&	 0.2759 	&	 0.8853(91)		&	 225\\
 5.85 	&	 8 	&	 0.693 	&	 0.2865 	&	 0.9263(274)		&	 400\\
 5.83 	&	 8 	&	 0.667 	&	 0.2977 	&	 0.9434(136)		&	 325\\
 5.80 	&	 8 	&	 0.629 	&	 0.3159 	&	 0.9619(112)		&	 402\\
 5.72 	&	 8 	&	 0.530 	&	 0.3744 	&	 0.9966(555)		&	 400\\
\hline
 6.33$^*$ 	&	 12 	&	 0.989 	&	 0.1338 	&	 0.4535(153)		&	 117\\
 6.32$^*$ 	&	 12 	&	 0.975 	&	 0.1356 	&	 0.4473(138)		&	 168\\
 6.31$^*$ 	&	 12 	&	 0.962 	&	 0.1375 	&	 0.5116(214)		&	 197\\
 6.30$^*$ 	&	 12 	&	 0.948 	&	 0.1395 	&	 0.5163(119)		&	 218\\
 6.29 	&	 12 	&	 0.935 	&	 0.1415 	&	 0.5647(40)		&	 225\\
 6.28 	&	 12 	&	 0.922 	&	 0.1435 	&	 0.6092(53)		&	 225\\
 6.27 	&	 12 	&	 0.909 	&	 0.1455 	&	 0.6633(44)		&	 225\\
 6.26 	&	 12 	&	 0.896 	&	 0.1476 	&	 0.6437(74)		&	 225\\
 6.25 	&	 12 	&	 0.883 	&	 0.1497 	&	 0.7346(34)		&	 225\\
 6.24 	&	 12 	&	 0.871 	&	 0.1519 	&	 0.6517(172)		&	 225\\
 6.22 	&	 12 	&	 0.846 	&	 0.1564 	&	 0.6959(134)		&	 223\\
 6.20 	&	 12 	&	 0.822 	&	 0.1610 	&	 0.7311(116)		&	 225\\
 6.16 	&	 12 	&	 0.774 	&	 0.1709 	&	 0.8207(65)		&	 275\\
 6.12 	&	 12 	&	 0.729 	&	 0.1815 	&	 0.8858(33)		&	 258\\
 6.08 	&	 12 	&	 0.685 	&	 0.1931 	&	 0.9258(125)		&	 325\\
 6.00 	&	 12 	&	 0.602 	&	 0.2197 	&	 0.9750(166)		&	 325\\
 5.90 	&	 12 	&	 0.507 	&	 0.2611 	&	 0.9876(180)		&	 425\\
\end{tabular}
\end{ruledtabular}
\end{table}

We generate SU(3) pure gauge lattice configurations in NVIDIA GPUs, Graphical Processing Unit, (GTX295, GTX480, GTX580 and Tesla C2070). 
The SU(3) CUDA code uses the standard Wilson action via combination of Cabibbo-Marinari pseudoheatbath and over-relaxation algorithm by three SU(2)-subgroups, \cite{Cabibbo:1982zn,Cardoso:2010di,ptqcd}.
In order to reduce memory traffic, we store only the first two rows of SU(3) matrices in the GPU Global memory and reconstruct the third line of each matrix on the fly when needed.
Each iteration consists of 4 pseudoheatbath and 7 over-relaxation steps followed link reunitarization. 
After the thermalization, in order to generate a Markov chain of configurations, we only store a configuration, to evaluate the colour averaged free energy, every 200 iterations.
In order to scan a large number of temperatures, and to verify our results, we consider 57 different sets of configurations, with 225 to 450 configurations each, for different $\beta$, $N_t$ and lattice sizes. In total we produced the large number of circa three million SU(3) configurations, thus reaching the limit of our computational facilities.

\begin{figure}[t!]
\begin{center}
    \includegraphics[width=8.5cm]{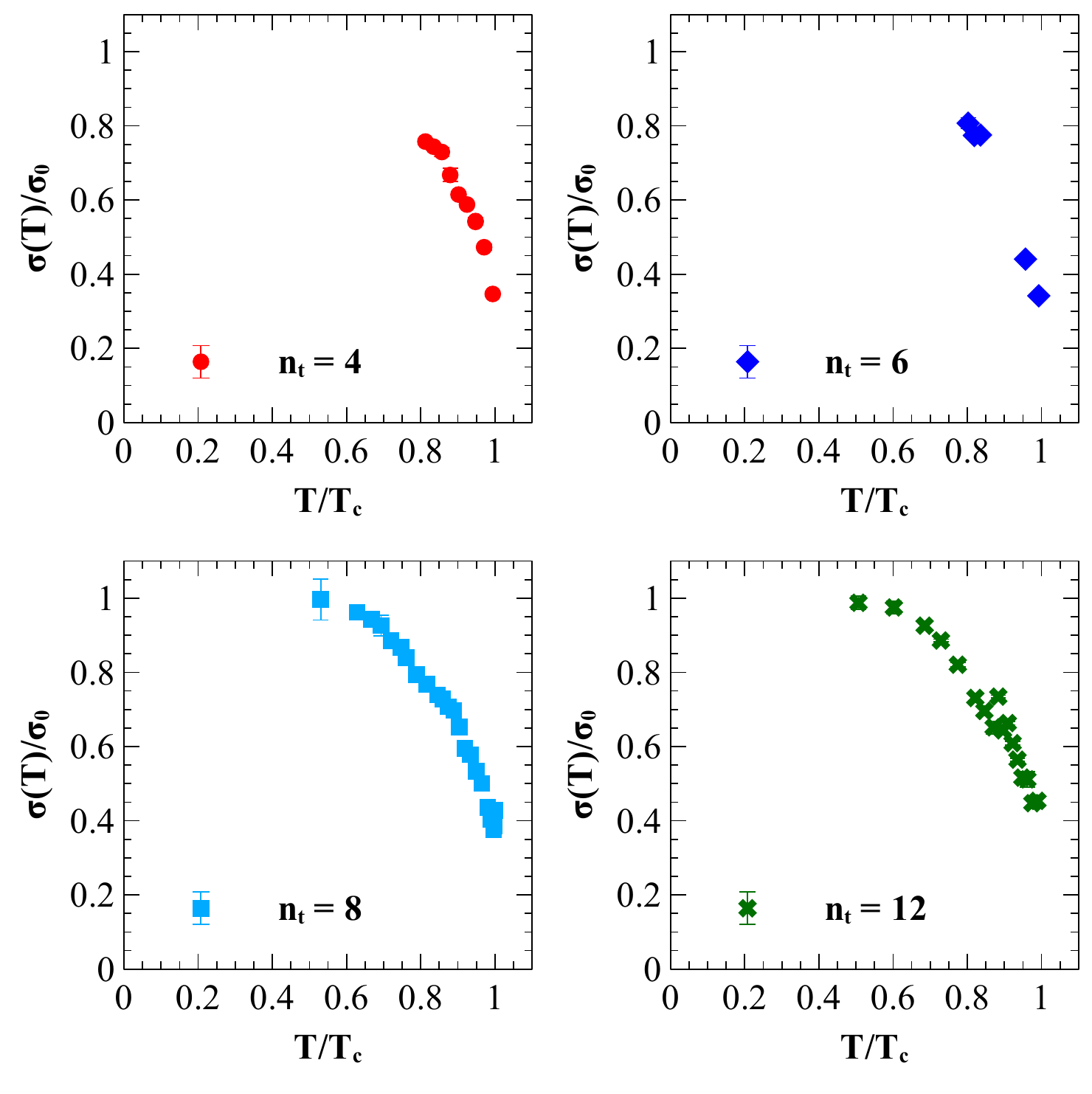}
    \caption{String tension as function of $T / T_c$ for four temporal extensions, $N_t=4,6,8,12$.}
    \label{string_allntall}
\end{center}
\end{figure}

\section{Results for $\sigma(T)$ and outlook }

We fit the colour averaged free energy with Eq. (\ref{ffit}), fixing the Coulomb term with the L\"{u}scher term. Notice that close to $T_c$ the linear term decreases, while the Coulomb term remains dominated by  the constant L\"uscher term, and we have go to distances as large as possible in order to to fit the string tension. Thus we fit the string tension in an interval at the largest distance before the statistical noise is large, and before the periodicity of the lattice saturates the free energy. 
The fitting range is illustrated in Fig. \ref{fig:Fit_48_3_8_b5.92}.
In all our fits, the distance interval is sufficient to have a $\chi^2/dof$ smaller than one.
The error in the mean average free energy, as well as the fit parameters, were determined by jackknife method.

Since large quark-antiquark distances are important, we also test the large volume limit for the configuration set with $\beta=5.67$, $N_t=4$ and $T=0.948 \, T_c$.  
We illustrate in Fig. \ref{stft_nt4_b5} four different spatial lattice volumes, $24^3$, $32^3$, $48^3$ and $64^3$ , in order to detect spatial volume dependence effects in the results. 
We respectively obtain the fits for $\sigma(T)/\sigma(0)$  of $0.462 \pm 0.012$, $0.475 \pm 0.023$, $0.541 \pm 0.006$ and $0.546 \pm 0.007$. 
For $N_t=48$ and $N_t=64$, the values obtained are already very close and within the statistical error.  
Since it would be too demanding in computer time to apply the same large volume limit extrapolation to all our lattice QCD configurations, 
we use the same spatial lattice volume of $48^3$ in lattice units.

In Table \ref{tab:stringtension} and in Fig. \ref{string_allntall} we show the results of our fits for the 54 configurations sets in $48^3 \times N_t$ lattices we generate at different $T$.
We utilize four different temporal extensions $N_t=4,6,8\text{ and }12$ because we want to scan a wide temperature range between $T=0$ and $T_c$.
We observe that at $T \simeq 0.5\,T_c$, the string tension is practically equal to the string tension at zero temperature $\sigma_0$.
Thus, to limit the study of lattices requiring a larger $N_t$, we address the interval $[0.5 T_c, T_c[$.

\begin{figure}[t!]
\begin{center}
    \includegraphics[width=7cm]{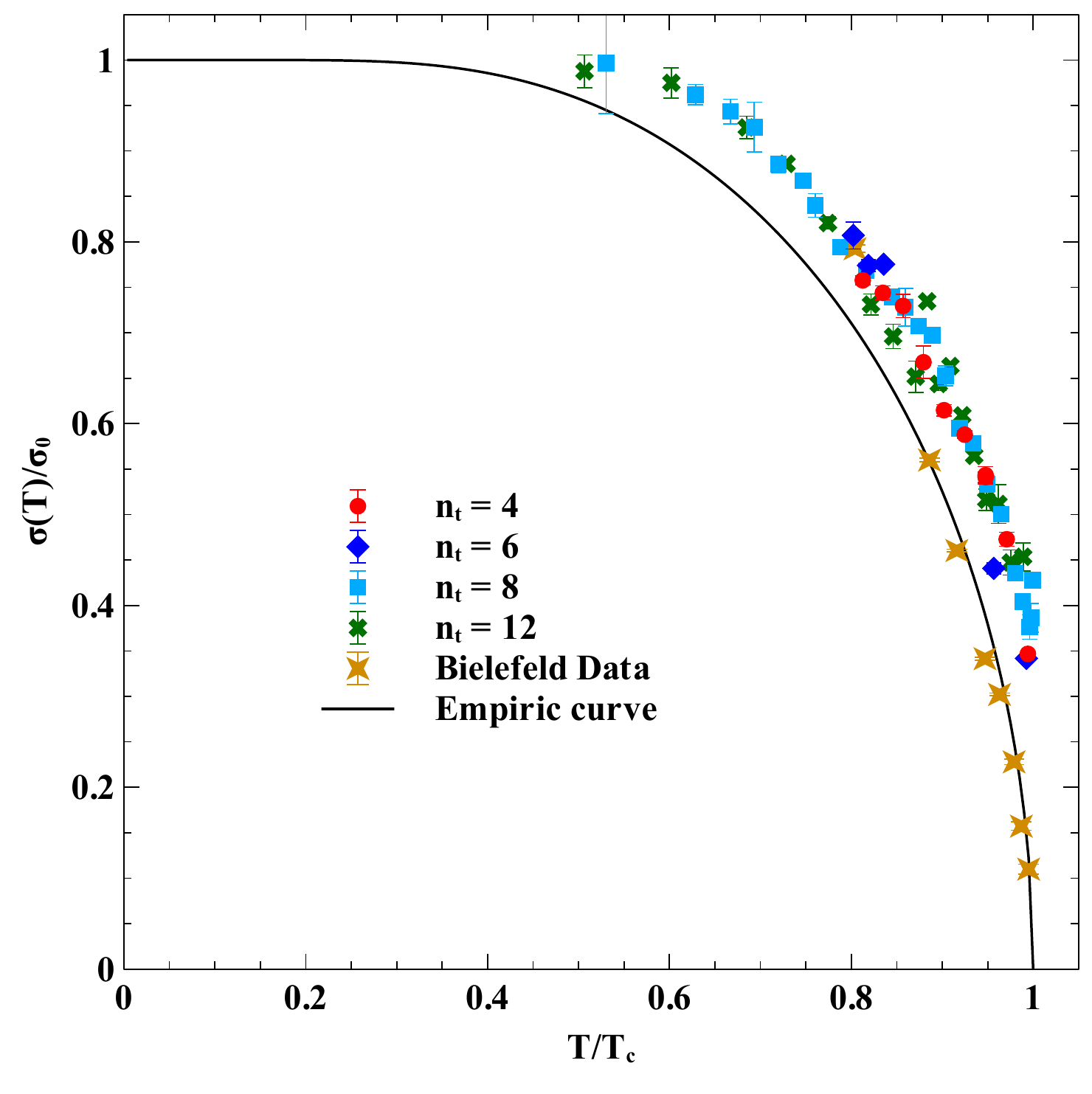}
    \caption{String tension as function of the temperature for $N_t=4,6,8,12$ combined with the results from the Bielefeld group. The black line corresponds to the ferromagnet magnetization $M/Msat$ critical curve, \cite{Bicudo:2010hg}.}
    \label{stringtensionall}
\end{center}
\end{figure}

In Fig. \ref{stringtensionall}, 
we compare the summary of our results with the results obtained by the Bielefeld group
\cite{Kaczmarek:1999mm}
 and the empirical curve \cite{Bicudo:2010hg}. 
The Bielefeld data was obtained from lattice gauge configurations with $32^3\times 4$ size and a tree level Symanzik-improved action. 
We utilize several $N_t$, the simpler Wilson action and a simpler fit with only a constant term, the L\"uscher Coulomb term and the linear confinement.
The empirical curve essentially fits most of the Bielefeld data, except for the lowest temperature one, the one at $T \simeq 0.8 T_c$. 
Using all the configurations we generated, our points would be above the curve up to higher temperatures circa $0.95 T_c$ and only beyond this temperature would get close to the empirical curve. 

However, since we have a finite volume, close to $T_c$ there is a contamination of deconfined configurations in our sets of configurations. Above $T \simeq 0.95 T_c$ We observe, utilizing the history of the Polyakov loop, a pair of gaussian-like peaks 
\cite{Kaczmarek:1999mm}
both in the norm and in the real part of a single Polyakov loop average. 
Since we have no computational power to further increase the volume, we remove the deconfined configurations from our Markov chain, with a cut between half way between the two gaussian peaks. The $\beta$ where this procedure was necessary are marked with an asterisk * in Table \ref{tab:stringtension}.
Without this deconfined configurations removal, some of the string tensions $\sigma({T_c}^-)$ would be as small as $0.1 \sigma_0$, consistent with the results of \cite{Kaczmarek:1999mm}. The points below $T \simeq 0.95 T_c$ together with the sets of confined only configurations at $T \geq 0.95 T_c $ indicate a first order discontinuity at $\sigma({T_c}^-) \simeq 0.4 \sigma_0$.

To clarify the extent of the deconfinement first order transition, we plan to address, in future works, the mapping of the $\sigma(T)$ with different lattice QCD techniques.

\vspace{-.0cm}
\acknowledgments
This work was partly funded by the FCT contracts, PTDC/FIS/100968/2008,  CERN/FP/109327/2009, CERN/FP/116383/2010 and 
 SFRH/BD/44416/2008.
We are very grateful to the Referee of Physical Review D and thank Orlando Oliveira for useful discussions.

\vspace{-.1cm}

\bibliographystyle{apsrev4-1}
\bibliography{bib}

\end{document}